\documentclass[preprint,showpacs,preprintnumbers,amsmath,amssymb]{revtex4}

\usepackage{graphicx}
\usepackage{dcolumn}
\usepackage{bm}

\begin{document}

\title{Dynamically generated resonances from the vector meson-octet baryon
interaction in the strangeness zero sector}

\author{Bao-Xi Sun}

\affiliation{College of Applied Sciences, Beijing University of
Technology, Beijing 100124, China}

\author{Xiao-Fu Lu}

\affiliation{Department of Physics, Sichuan University, Chengdu
610065, China}

\begin{abstract}
The interaction potentials between vector mesons and octet baryons
are calculated explicitly with a summation of $t-$, $s-$, $u-$
channel diagrams and a contact term originating from the tensor
interaction.
Many resonances are generated dynamically in different channels of
strangeness zero by solving the coupled-channel Lippman-Schwinger
equations with the method of partial wave analysis, and their total
angular momenta are determined.
The spin partners $N(1650)1/2^{-}$ and $N(1700)3/2^-$,
$N(1895)1/2^{-}$ and $N(1875)3/2^-$, and the state $N(2120)3/2^-$
are all produced respectively in the isospin $I=1/2$ sector. In the
isospin $I=3/2$ sector, the spin partners $\Delta(1620)1/2^-$ and
$\Delta(1700)3/2^-$ are also associated with the pole in the complex
energy plane. According to the calculation results, a $J^P=1/2^-$
state around 2000 MeV is predicted as the spin partner of
$N(2120)3/2^-$.
Some resonances are well fitted with their counterparts listed in
the newest review of Particle Data Group(PDG)\cite{pdg}, while
others might stimulate the experimental observation in these energy
regions in the future.
\end{abstract}


\maketitle

\section{Introduction}

The combination of chiral Lagrangian with nonperturbative unitary
techniques in coupled channels of mesons and baryons has become a
powerful method to study the meson-meson and meson-baryon
interactions and new states in the resonance region, which are not
easily explained using the conventional constituent quark model. The
analysis of the meson baryon scattering matrices shows poles in the
second Riemann sheet that can be associated to known resonances or
new ones. In this way, $J^P=1/2^-$ resonances have been generated
dynamically in the interaction of pseudoscalar mesons with octet
baryons, which fit quite well the spectrum of the known low-lying
resonances with these quantum numbers\cite{Kaiser1995, weise,
Kaiser1996, angels, ollerulf, carmina, carmenjuan, hyodo,
Hyodo2006}. Similarly,  the $J^P=3/2^-$ resonances are obtained in
the interaction of pseudoscalar mesons with decuplet
baryons\cite{Lutz, sarkar}.

The theory on the hidden gauge symmetry supplies a mechanism to
include vector mesons in the chiral
Lagrangian\cite{hidden1,hidden2,hidden3,hidden4}. Therefore, the
study on the interaction between vector mesons and other hadrons
becomes possible. Along this clue, the interactions between vector
mesons and psuedoscalar mesons are studied and the radiative decays
of some axial-vector mesons generated dynamically are
discussed\cite{Nagahiro}. Similarly, the scattering amplitudes of
two $\rho$ mesons are calculated within the framework of
coupled-channel Lippman-Schwinger equations, and two resonances
$f_0(1370)$ and $f_2(1270)$ are generated dynamically\cite{Molina}.
This work has been extended to the $SU(3)$-space of vector mesons in
Ref.~\cite{Geng}, where several known resonances are also
dynamically generated.
In the baryon sector, the interaction of vector mesons and decuplet
baryons is addressed in Refs.~\cite{Gonzalez,souravbao}, where only
$t-$ channel amplitudes are analyzed in the S-wave approximation.
This method is also extended to study the interaction of vector
mesons and octet baryons, and several baryon resonances have been
found as a result of solving the coupled-channel Lippman-Schwinger
equations\cite{ramos2010,Garzon2012,Garzon2013}. However, the
resonances generated dynamically are spin degenerate since the
amplitude obtained from $t-$ channel interaction is independent of
spin.
Because the masses of vector mesons are comparable to those of
baryons, only $t-$ channel diagrams might be incomplete to obtain a
reliable interaction of vector mesons and baryons. Thus in
Ref.~\cite{Hosaka2011}, the $t-$,$s-$, $u-$ channel diagrams and a
contact diagram originating from the tensor term of the vector
meson-octet baryon interaction are all taken into account, and four
spin-determined resonances are found in a non-relativistic
approximation of the coupled-channel Lippman-Schwinger equations.
In the present work, we deduce the interaction kernel of vector
mesons and octet baryons including a vector meson exchange in $t-$
channel, octet baryon exchange in $s-$, $u-$ channels, and a contact
diagram related only to the tensor interaction term in a fully
relativistic framework, and then calculate the scattering amplitudes
of vector mesons and octet baryons by solving the coupled-channel
Lippman-Schwinger equations. The amplitude of the vector mesons and
octet baryons will be expanded in terms of partial waves, and then
the poles of the amplitudes in different partial waves are detected
in the complex energy plane in center of mass system, which can be
associated to some well-known resonances.

In Sect.~\ref{sect:Formalism}, we will show the basic Lagrangian
obtained with the hidden gauge symmetry of $SU(3)$ group, where a
tensor interaction term is included. Then the framework on the
coupled-channel Lippman-Schwinger equations will be summarized
briefly. In Sect.~\ref{sect:Parameters}, the parameters in the
Lagrangian which are fitted with the experimental data on coupling
constants of octet baryons to vector mesons are determined.
In Sect.~\ref{sect:pwa}, the amplitudes are expanded in terms of
partial waves, and the formula on the partial wave analysis is
displayed.
In Sect.~\ref{sect:Results}, we will analyze the resonances found in
the complex energy plane for the vector meson-octet baryon system
with total strangeness zero, and the properties of these resonances
and their possible PDG counterparts are discussed. Finally, we will
present a summary on this article.

\section{Formalism}
\label{sect:Formalism}

We follow the formalism of the hidden gauge interaction for vector mesons of
\cite{hidden1,hidden2,hidden3,hidden4} in this manuscript. The Lagrangian involving the interaction of vector mesons among themselves is given by
\begin{equation}
\label{eq:3vector}
\mathcal{L}_{V} ~=~ - \frac{1}{2} \langle V^{\mu\nu} V_{\mu\nu} \rangle,
\end{equation}
where the symbol $\langle \rangle$ stands for the trace in the $SU(3)$ space and the tensor field of vector mesons is given by
\begin{equation}
\label{eq:vector tensor}
V^{\mu\nu} = \partial^{\mu} V^\nu - \partial^{\nu} V^\mu - ig \left[V^\mu, V^\nu \right],
\end{equation}
where $g$ is
\begin{equation}
g~=~\frac{M_V}{\sqrt{2} f_\pi},
\end{equation}
with $f_\pi~=~93$MeV the pion decay constant and $M_V$ the mass of
the $\rho$ meson. The magnitude $V_\mu$ is defined by the matrix
\begin{eqnarray}
V_\mu =\frac{1}{2} \left( \begin{array}{ccc}
\rho^0 + \omega & \sqrt{2}\rho^+ & \sqrt{2}K^{*^+}\\
&& \\
\sqrt{2}\rho^-& -\rho^0 + \omega & \sqrt{2}K^{*^0}\\
&&\\
\sqrt{2}K^{*^-} &\sqrt{2}\bar{K}^{*^0} & \sqrt{2} \phi
\end{array}\right)_\mu.
\end{eqnarray}
The interaction of ${\cal L}_{V}$ gives rise to a three-vector vertex form
\begin{equation}
\label{eq:3V}
{\cal L}_{(3V)}=i2g\langle (\partial_\mu V_\nu -\partial_\nu V_\mu) V^\mu V^\nu\rangle
,
\end{equation}
which will contribute to the $t-$ channel of the vector meson-octet
baryon interaction.

The tensor interaction term can be included in the vector
meson-octet baryon Lagrangian as is done in Ref.~\cite{Hosaka2011}.
It is no doubt that the tensor term also satisfies the $SU(3)$
hidden gauge symmetry and might make an amendment to the vector
meson-octet baryon interaction. Therefore, the lagrangian for the
vector meson-octet baryon interaction due to the SU(3) hidden gauge
symmetry can be written as
\begin{eqnarray} \label{eq:vbb}
\mathcal{L}_{VB}&=& -g \Biggl\{ F_V \langle \bar{B} \gamma_\mu \left[ V^\mu, B \right] \rangle + D_V \langle \bar{B} \gamma_\mu \left\{ V^\mu, B \right\} \rangle + \langle \bar{B} \gamma_\mu B \rangle  \langle  V^\mu \rangle
\Biggr. \\ \nonumber
&+&\left. \frac{1}{4 M} \left( F_T \langle \bar{B} \sigma_{\mu\nu} \left[ V^{\mu\nu}, B \right] \rangle  + D_T \langle \bar{B} \sigma_{\mu\nu} \left\{ V^{\mu\nu}, B \right\} \rangle\right)\right\},
\end{eqnarray}
where $B$ is the $SU(3)$ matrix of octet baryons
\begin{eqnarray}
B =
\left( \begin{array}{ccc}
 \frac{1}{\sqrt{6}} \Lambda + \frac{1}{\sqrt{2}} \Sigma^0& \Sigma^+ & p\\
&& \\
\Sigma^-&\frac{1}{\sqrt{6}} \Lambda- \frac{1}{\sqrt{2}} \Sigma^0 &n\\
&&\\
\Xi^- &\Xi^0 & -\sqrt{\frac{2}{3}} \Lambda
\end{array}\right),
\end{eqnarray}
and $M$ is the mass of the nucleon.

In order to obtain the correct couplings to the physical $\omega$
and $\phi$ meson, the mixing of their octet and singlet components
must be considered. Under the ideal mixing assumption, we have
\begin{eqnarray}
\omega&=&\sqrt{1/3}\omega_8~+~\sqrt{2/3}\omega_0, \nonumber \\
\phi&=&-\sqrt{2/3}\omega_8~+~\sqrt{1/3}\omega_0, \nonumber \\
\end{eqnarray}
and only the octet parts of these wave functions are included in
Eq.~(\ref{eq:vbb}). For the singlet states we have
\begin{eqnarray}
\label{eq:singlet}
\mathcal{L}_{V_0BB} = -g \Biggl\{\frac{C_V}{3}  \langle \bar{B} \gamma_\mu B \rangle  \langle  V_0^\mu \rangle
+ \frac{ C_T}{4 M}  \langle \bar{B} \sigma_{\mu\nu}  V_0^{\mu\nu} B  \rangle  \Biggl\}.
\end{eqnarray}
Thus, the Lagrangian for the linearly coupling to vector mesons can
be explicitly written as
\begin{eqnarray}
\label{yukawaL}
\mathcal{L}_{VBB}&=& -g \biggl\{ F_V \langle \bar{B} \gamma_\mu \left[ V_8^\mu, B \right] \rangle
+D_V \langle \bar{B} \gamma_\mu \left\{ V_8^\mu, B \right\} \rangle \nonumber \\
&&+ \frac{1}{4 M} \Bigl( F_T \langle \bar{B} \sigma_{\mu\nu} \left[ \partial^{\mu} V_8^\nu - \partial^{\nu} V_8^\mu, B \right] \rangle \Bigr.
\biggr.
+\Bigl.  D_T \langle \bar{B} \sigma_{\mu\nu} \left\{ \partial^{\mu} V_8^\nu - \partial^{\nu} V_8^\mu, B \right\} \rangle\Bigr) \\ \nonumber
&+& \frac{C_V}{3} \langle \bar{B} \gamma_\mu B \rangle  \langle  V_0^\mu \rangle
+\left.  \frac{ C_T}{4 M}  \langle \bar{B} \sigma_{\mu\nu}  V_0^{\mu\nu} B  \rangle\right\},
\end{eqnarray}
which will contribute to the $t-$, $s-$, $u-$ channel interactions
between vector mesons and octet baryons.
Moreover, the self-coupling terms of vector mesons in
Eq.~(\ref{eq:vector tensor}) lead to a contact interaction between
vector mesons and octet baryons, which is trivially null for the
singlet vector meson, thus
\begin{equation}
\label{contactl}
\mathcal{L}_{VVBB} =  \frac{g}{4 M} \Bigl\{ F_T \langle \bar{B} \sigma_{\mu\nu} \left[ ig \left[V_8^\mu, V_8^\nu \right], B \right] \rangle  + D_T \langle \bar{B} \sigma_{\mu\nu} \left\{  ig \left[V_8^\mu, V_8^\nu \right], B \right\} \rangle \Bigr\}.
\end{equation}

From Eqs.~(\ref{eq:3V}), (\ref{yukawaL}) and (\ref{contactl}), we
can obtain the potentials for the $t-$, $s-$, $u-$ channel and
contact interactions between vector mesons and octet baryons. If the
momentum of the initial vector meson is similar to that of the final
vector meson, i.e.,$q_2~\approx~q_1$, the momentum transfer
$k=q_2-q_1$ is trivial null approximately, and then the $t-$ channel
interaction can be written as
\begin{eqnarray}
\label{eqt-ch}
V^t_{ij}&=&-\frac{g}{\mu^2}
\left( \bar{U}(p_2,\lambda_2) \Gamma_\mu(p_2,p_1) U(p_1, \lambda_1) (q_1^\mu + q_2^\mu) \right) \varepsilon(q_1, \delta_1) \cdot \varepsilon^\ast(q_2, \delta_2),
\end{eqnarray}
where
\begin{equation}
\Gamma^\mu(p_2,p_1)~=~g_1 \gamma^\mu~+~g_2 (p_2^\mu~+~p_1^\mu)
\end{equation}
is the vertex of two baryons and a vector meson when the tensor term
is taken into account in the Lagrangian of Eq.~(\ref{eq:vbb}), and
the coupling constants $g_1$ and $g_2$ for different octet baryons
and vector mesons are attached in Appendix I. In the above equation,
$U(p_1, \lambda_1)$ and $\bar{U}(p_2,\lambda_2)$ are the wave
functions of the incoming and outgoing baryons, and
$\varepsilon(q_1, \delta_1)$ and $\varepsilon^\ast(q_2, \delta_2)$
are polarization vectors of the initial and final mesons,
respectively\cite{Lurie}. The formulas of them can be found in
Appendix II.
However, if the difference between $q_2$ and $q_1$ is taken into
account, an additional part in Eq.~(\ref{eqt-supp}) must be
supplemented in the $t-$ channel interaction of vector mesons and
octet baryons.
\begin{eqnarray}
\label{eqt-supp}
V^t_{supp,ij}&=&\frac{2g}{\mu^2}\left(
\left\{ \bar{U}(p_2,\lambda_2) \Gamma^\mu(p_2,p_1) \varepsilon^\ast_\mu(q_2, \delta_2) U(p_1, \lambda_1)  \right\} q_2 \cdot \varepsilon(q_1, \delta_1) \right.\\ \nonumber
&+& \left.  \left\{ \bar{U}(p_2,\lambda_2) \Gamma_\mu(p_2,p_1) \varepsilon^\mu(q_1, \delta_1) U(p_1, \lambda_1)  \right\} q_1 \cdot \varepsilon^\ast(q_2, \delta_2)  \right).
\end{eqnarray}

\begin{figure}[htb]
\begin{center}
\includegraphics[width=0.7\textwidth]{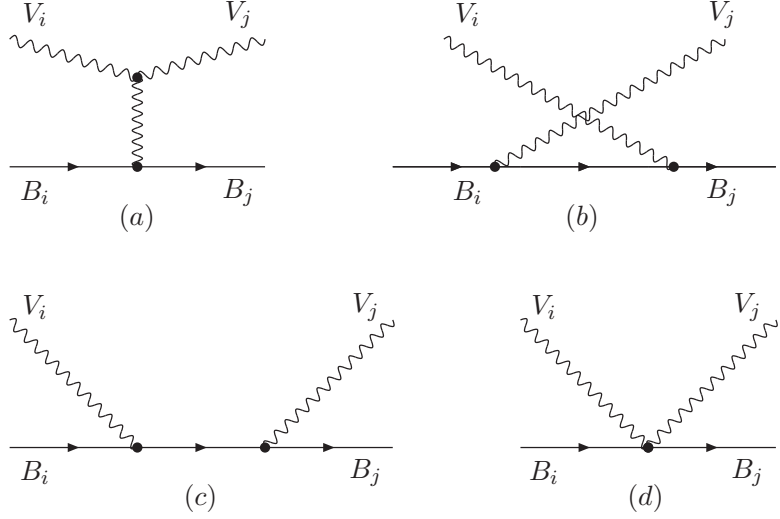}
\end{center}
\caption{Feynman diagrams of the vector meson-baryon interaction.
$(a)$ $t-$ channel, $(b)$ $u-$ channel, $(c)$ $s-$ channel, and
$(d)$ contact term.} \label{fig:channel}
\end{figure}

In addition to the $t-$ channel mechanism, the $u-$ channel and $s-$
channel mechanisms depicted in Fig.~\ref{fig:channel} are also
considered in this work. The $s-$ channel interaction of vector
mesons and octet baryons can be written as
\begin{equation}
\label{eqs-ch} V^s_{ij}~=~\bar{U}(p_2,\lambda_2)
\Gamma^\mu(p_2,p_1+q_1) \varepsilon^\ast_\mu(q_2,
\delta_2)\frac{\rlap{/}p_1+\rlap{/}q_1+M}{s-M^2}
\Gamma^\nu(p_1+q_1,p_1) \varepsilon_\nu(q_1, \delta_1) U(p_1,
\lambda_1)
\end{equation}
with $s=(p_1+q_1)^2$, while the $u-$ channel interaction is
\begin{equation}
\label{equ-ch}
 V^u_{ij}~=~\bar{U}(p_2,\lambda_2)
\Gamma^\mu(p_2,p_1-q_2) \varepsilon_\mu(q_1,
\delta_1)\frac{\rlap{/}p_1-\rlap{/}q_2+M}{(p_1-q_2)^2-M^2}
\Gamma^\nu(p_1,p_1-q_2)\varepsilon^\ast_\nu(q_2, \delta_2) U(p_1,
\lambda_1).
\end{equation}
Since the three-momenta of vector mesons and octet baryons are far
smaller than their masses in the concerned energy region, we make an
approximation of $(p_1-q_2)^2\approx M_1^2+m_2^2-2 M_1 m_2$ in the
propagator in Eq.~(\ref{equ-ch}), where $M_1$ and $m_2$ are the
masses of initial octet baryons and final vector mesons,
respectively.

From Eq.~(\ref{contactl}), the contact interaction of vector mesons
and octet baryons is obtained
\begin{equation}
\label{eqCT} V^{CT}_{ij}~=~-i C^{CT}_{IS} 2 \bar{U}(p_2,\lambda_2)
\varepsilon^\ast_\mu(q_2, \delta_2) \sigma^{\mu\nu}
\varepsilon_\nu(q_1, \delta_1)  U(p_1, \lambda_1).
\end{equation}

Altogether, the total kernel of the vector menson-octet baryon
interaction can be written as
\begin{equation}
\label{eq:kernel} V_{ij}(s,t) = V^t_{ij} + V^t_{supp,ij} + V^s_{ij}
+ V^u_{ij} + V^{CT}_{ij},
\end{equation}
which is a summation of the $t-$, $s-$, $u-$ channels and contact
interaction. Now the kernel in Eq.~(\ref{eq:kernel}) is a function
of the total energy in the center of mass system $\sqrt{s}$ and the
scattering angle $\theta$.

The coefficients in Eqs.~(\ref{eqt-ch}), (\ref{eqt-supp}),
(\ref{eqs-ch}),(\ref{equ-ch}) and (\ref{eqCT}) can be obtained both
in the physical basis of states or in the isospin basis, and this is
discussed in Appendix I. In this article, we will directly study the
interaction in the isospin basis. Especially, we will concentrate on
the states of isospin $I=\frac{1}{2}$ and $\frac{3}{2}$ with
strangeness zero.

The scattering matrix implies solving the coupled-channel
Lippman-Schwinger equations in the on-shell factorization approach
of\cite{angels,ollerulf}
   \begin{equation}
T(\sqrt{s},\cos{\theta}) = [1 - V(\sqrt{s},\cos{\theta}) \,
G(s)]^{-1}\, V(\sqrt{s},\cos{\theta}) \label{eq:Bethe}
\end{equation}
with $G(s)$ being the loop function of a vector meson and a baryon
which we calculate in dimensional regularization using the formula
of \cite{ollerulf}
\begin{eqnarray}
G_{l}(s) &=& i 2 M_l \int \frac{d^4 q}{(2 \pi)^4} \,
\frac{1}{(P-q)^2 - M_l^2 + i \epsilon} \, \frac{1}{q^2 - m^2_l + i
\epsilon} \nonumber
\\ &=& \frac{2 M_l}{16 \pi^2} \left\{ a_l(\mu) + \ln
\frac{M_l^2}{\mu^2} + \frac{m_l^2-M_l^2 + s}{2s} \ln
\frac{m_l^2}{M_l^2} + \right. \nonumber \\ & &  \phantom{\frac{2
M}{16 \pi^2}} + \frac{\bar{q}_l}{\sqrt{s}} \left[
\ln(s-(M_l^2-m_l^2)+2\bar{q}_l\sqrt{s})+
\ln(s+(M_l^2-m_l^2)+2\bar{q}_l\sqrt{s}) \right. \nonumber  \\
& & \left. \phantom{\frac{2 M}{16 \pi^2} +
\frac{\bar{q}_l}{\sqrt{s}}} \left. \hspace*{-0.3cm}-
\ln(-s+(M_l^2-m_l^2)+2\bar{q}_l\sqrt{s})-
\ln(-s-(M_l^2-m_l^2)+2\bar{q}_l\sqrt{s}) \right] \right\} \ ,
\label{eq:gpropdr}
\end{eqnarray}
with $\mu$ a regularization scale, which we take to be 630 MeV, and
with a natural value of the subtraction constant $a_l(\mu)$ of $-2$,
as determined in \cite{ollerulf,ramos2010}.

In eq.~(\ref{eq:gpropdr}), $\bar{q}_l$ denotes the three-momentum of
the vector meson or the octet baryon in the center of mass frame and
is given by
\begin{equation}
\bar{q}_l=\frac{\lambda^{1/2}(s,m_l^2,M_l^2)}{2\sqrt{s}}
=\frac{\sqrt{s-(M_l+m_l)^2}\sqrt{s-(M_l-m_l)^2}}{2\sqrt{s}},
\end{equation}
where $\lambda$ is the triangular function and $M_l$ and $m_l$ are
the masses of octet baryons and vector mesons, respectively.

In order to find the pole corresponding to the resonance, we must
extend our calculation of the $T$ matrix to the complex plane of
$\sqrt{s}$. Because the physical G-propagators in Eq.~
(\ref{eq:gpropdr}) have cut-lines in the real axis, they are valid
only in the first Riemann sheet. However, the poles corresponding
resonances can only be found in the second Riemann sheet, and the
G-propagators in the second Riemann sheet must be studied.

The G-propagator near the real axis in the second Riemann sheet is
defined as
\begin{equation}
\label{eq:G-prop}
  G_l^{II}(s) =
 \left\{
 \begin{array}{lcc}
  G_l(s)   &
  ~ \mbox{at} & \mbox{Re}[\sqrt{s}] \le \sqrt{s^0}_l
  \\
  G_l(s) - 2 i \mbox{Im}[G_l(s)] &
  ~ \mbox{at} & \mbox{Re}[\sqrt{s}]  >  \sqrt{s^0}_l
 \end{array}
 \right\},
\end{equation}
with $\sqrt{s^0}_l=M_l+m_l$ the $l$-th channel threshold energy .

According to Eq. (\ref{eq:gpropdr}), the imaginary part of the
G-propagator in the dimensional regularization scheme can be written
as
\begin{equation}
\mbox{Im}[G_l^{I}(s)]=-\frac{M_l q_l}{4\pi\sqrt{s}}.
\label{eq:imagg}
\end{equation}

Since the widths of the $K^*$ and $\rho$ mesons are considerably
large, their mass distributions must be taken into account. We
follow the traditional method of convoluting the $G$-function with
the mass distributions of the $K^*$ and $\rho$ mesons, as is used in
Ref.~\cite{ramos2010}.
\begin{eqnarray}
\tilde{G}(s)=
\frac{1}{N}\int^{(m+2\Gamma_i)^2}_{(m-2\Gamma_i)^2}d\tilde{m}^2
\left(-\frac{1}{\pi}\right) {\rm Im}\,\frac{1}{\tilde{m}^2-m^2+{\rm
i} \tilde{m} \Gamma(\tilde{m})} & G(s,\tilde{m}^2,\tilde{M}^2_B)\ ,
\label{Gconvolution}
\end{eqnarray}
where the normalization factor is written as
\begin{equation}
N=\int^{(m_\rho+2\Gamma_i)^2}_{(m_\rho-2\Gamma_i)^2}d\tilde{m}^2
\left(-\frac{1}{\pi}\right){\rm
Im}\,\frac{1}{\tilde{m}^2-m^2_\rho+{\rm i} \tilde{m}
\Gamma(\tilde{m})} \label{Norm}
\end{equation}
with
\begin{equation}
\Gamma(\tilde{m}) = \Gamma_{i} \left(\dfrac{m^2}{\tilde{m}^2}\right)
\left(\dfrac{\lambda^{1/2}(\tilde{m}^2, m_d^2, m_d^{\prime\,
2})/2\tilde{m}} {\lambda^{1/2}(m^2, m_d^2, m_d^{\prime\,
2})/2m}\right)^3 \theta \left( \tilde{m} - m_d - m_d^\prime \right),
\label{eq:Gamma}
\end{equation}
and $\Gamma_i$ the decay width of the meson ($i=\rho,K^*$), which we
take to be 149.1 MeV and 49.75 MeV for the $\rho$ and  $K^*$ meson,
respectively. In Eq.~(\ref{eq:Gamma}), $m_d, m_d^\prime$ denote the
masses of the decay products of the vector mesons,  i.e., pion
masses in case of $\rho$ for the decay mode of $\rho \rightarrow \pi
\pi$, and kaon and pion masses in case of $K^*$ for $K^* \rightarrow K
\pi$.

As far as the mass distributions of the vector mesons are concerned,
the $G$-function in the coupled-channel Lippman-Schwinger equations
(\ref{eq:Bethe}) should be replaced by $\tilde{G}(s)$ in
Eq.~(\ref{Gconvolution}).

\section{Parameters}
\label{sect:Parameters}

\begin{table}[hbt]
\begin{center}
\begin{tabular}{c|cccc}
\hline
 $Exp.$    & $\omega$ & $\rho$ & $K^*$  & $\phi$  \\
\hline
$g^2/4\pi$ &    2.4   &    2.4 & 1.39   & 12.0     \\
\hline
\end{tabular}
\caption{The experimental values on the coupling constants of vector
mesons to octet baryons\cite{Qcdsumrule}.  } \label{tabcoupling}
\end{center}
\end{table}

\begin{table}[hbt]
\begin{center}
\begin{tabular}{cccccc}
\hline
 $F_V$ & $D_V$ & $F_T$ & $D_T$ & $C_V$ & $C_T$ \\
\hline
 1.6405 & 0.2225 & 1.6405 & 0.2225 & -5.144 & -5.144 \\
\hline
\end{tabular}
\caption{The parameters used in the calculation.}
\label{tabparameter}
\end{center}
\end{table}

The experimental values on coupling constants of vector mesons to
octet baryons are listed in Table~\ref{tabcoupling}, which are taken
from Ref.~\cite{Qcdsumrule}. Thus we can fit the parameters $F_V$,
$D_V$ and $C_V$ in the Lagrangian with these data.
There is no doubt that the tensor terms are related to the form
factors of baryons, which can be deduced from the linear couplings
of vector mesons to octet baryons. However, the contribution from
tensor terms to the kernels at tree diagram levels might increase
the precision of the calculation.
Therefore, we set the parameters $F_T$, $D_T$ and $C_T$ related to the
tensor terms to be equal to $F_V$, $D_V$ and $C_V$, respectively.
The values of these parameters are listed in Table~\ref{tabparameter}.

\section{Partial wave analysis}
\label{sect:pwa}

If we set $l$, $j$ and $\mu$ the orbital angular momentum, the total
angular momentum and the orientation of total angular momentum of
the initial vector meson, $l^\prime$, $j^\prime$ and $\mu^\prime$
those of the final vector meson, and $J$ the total angular momentum
of the system, the scattering amplitudes of vector mesons and octet
baryons can be expanded in terms of partial waves. We have
\begin{eqnarray}
\label{eq:pwa-1} &&< \rho^\prime, \nu^\prime|T|\rho,\nu>
~=~<\hat{\vec{q}}_2; 1, 1/2; \rho^\prime, \nu^\prime|T| \hat{\vec{q}}_1; 1, 1/2; \rho,\nu>  \nonumber \\
&=&\sum_{j, j^\prime, \mu, \mu^\prime, J}<\hat{\vec{q}}_2; 1, 1/2;
\rho^\prime, \nu^\prime| j^\prime, 1/2; \mu^\prime, \nu^\prime
> < j^\prime, 1/2; \mu^\prime, \nu^\prime |
j^\prime, 1/2; J, M > \\
&& < j^\prime, 1/2; J, M|T|j, 1/2; J, M
> <j, 1/2; J, M|j, 1/2; \mu, \nu > < j, 1/2; \mu, \nu
|\hat{\vec{q}}_1; 1, 1/2; \rho,\nu >,  \nonumber
\end{eqnarray}
with
\begin{eqnarray}
<\hat{\vec{q}}_2; 1, 1/2; \rho^\prime, \nu^\prime| j^\prime,
1/2; \mu^\prime, \nu^\prime
>&=&\sum_{l^\prime}Y_{l^\prime,
\mu^\prime-\rho^\prime}\left( \hat{\vec{q}}_2 \right) <l^\prime, 1;
\mu^\prime-\rho^\prime, \rho^\prime | j^\prime, \mu^\prime >
\end{eqnarray}
and
\begin{eqnarray}
< j, 1/2; \mu, \nu |\hat{\vec{q}}_1; 1, 1/2; \rho,\nu >&=&\sum_{l}
<l, 1; \mu-\rho, \rho| j, \mu> Y^*_{l,\mu-\rho} \left(
\hat{\vec{q}}_1 \right).
\end{eqnarray}
In Eq.~(\ref{eq:pwa-1}), $\rho$ and $\rho^\prime$ denote the
orientations of spins for the inial and final vector mesons, and
$\nu$ and $\nu^\prime$ the orientations of spins for the inial and
final baryons, respectively.

The spherical harmonics $Y^*_{l,\mu-\rho} \left( \hat{\vec{q}}_1
\right)$ may be simplified by choosing the axis of quantization
along the momentum of the initial vector meson $\vec{q}_1$, and then
$Y^*_{l,\mu-\rho} \left( \hat{\vec{q}}_1 \right)=\delta_{\mu,\rho}
[(2l+1)/4\pi]^{1/2}$.
In the S-wave approximation of partial wave analysis, only the final
state with the orbital angular momentum $l^\prime=0$ is studied.
Since the parity is conserved, the contribution from the initial
state with the orbital angular momentum $l=0,2$ must be taken into
account. Thus the scattering amplitudes can be expanded in terms of
the total angular momentum $J$ of the system, the orbital angular
momentum $l$ and the total angular momentum $j$ of the initial
vector meson.
\begin{eqnarray}
\label{eq:pwa}
&&< \rho^\prime, \nu^\prime|T|\rho,\nu> \nonumber \\
&=& \sum_{l, j, J,
M} <1, 1/2; \rho^\prime, \nu^\prime|J,M > T^J_{1,j;0,l} < J,M|j,
1/2, \rho, \nu
> < l,1; 0, \rho | j, \rho > \left(
\frac{2l+1}{4\pi} \right)^{1/2} \nonumber  \\
&=&
 < 1, 1/2; \rho^\prime, \nu^\prime | 1/2, \rho^\prime+\nu^\prime>
T^{J=1/2}_{1,1;0,0}
 < 1, 1/2; \rho, \nu | 1/2, \rho+\nu> \left(
\frac{1}{4\pi} \right)^{1/2} \nonumber  \\
&+&
 < 1, 1/2; \rho^\prime, \nu^\prime | 3/2, \rho^\prime+\nu^\prime>
T^{J=3/2}_{1,1;0,0}
 < 1, 1/2; \rho, \nu | 3/2, \rho+\nu> \left(
\frac{1}{4\pi} \right)^{1/2} \nonumber  \\
&+&
 < 1, 1/2; \rho^\prime, \nu^\prime | 1/2, \rho^\prime+\nu^\prime>
T^{J=1/2}_{1,1;0,2}
 < 1, 1/2; \rho, \nu | 1/2, \rho+\nu>  <2,1;0,\rho|1,\rho> \left(
\frac{5}{4\pi} \right)^{1/2} \nonumber  \\
&+&
 < 1, 1/2; \rho^\prime, \nu^\prime | 3/2, \rho^\prime+\nu^\prime>
T^{J=3/2}_{1,1;0,2}
 < 1, 1/2; \rho, \nu | 3/2, \rho+\nu>  <2,1;0,\rho|1,\rho> \left(
\frac{5}{4\pi} \right)^{1/2} \nonumber  \\
&+&
 < 1, 1/2; \rho^\prime, \nu^\prime | 3/2, \rho^\prime+\nu^\prime>
T^{J=3/2}_{1,2;0,2}
 < 2, 1/2; \rho, \nu | 3/2, \rho+\nu>  <2,1;0,\rho|2,\rho> \left(
\frac{5}{4\pi} \right)^{1/2}, \nonumber  \\
\end{eqnarray}
with $<...|... >$ Clebsch-Gordan coefficients.

In Eq.~(\ref{eq:pwa}), the first and second terms are related to the
S-wave part of the initial vector meson, i.e., the case of $l=0$,
while the other terms correspond to the contributions from the case
of $l=2$, the D-wave part of the initial vector meson.

It is apparent that five amplitudes with different spin states of
the initial and final vector mesons and octet baryons need to be
calculated in order to obtain the values of amplitudes
$T^{J=1/2}_{1,1;0,0}$, $T^{J=3/2}_{1,1;0,0}$, $T^{J=1/2}_{1,1;0,2}$,
$T^{J=3/2}_{1,1;0,2}$ and $T^{J=3/2}_{1,2;0,2}$. We choose the
amplitudes $<1,1/2|T|1,1/2>$, $<0,1/2|T|0,1/2>$,
$<-1,1/2|T|-1,1/2>$, $<1,-1/2|T|0,1/2>$ and $<0,-1/2|T|-1,1/2>$ to
obtain these values when the spin symmetry is taken into account.

We search for poles of these amplitudes at the complex plane of
$\sqrt{s}$. If the $Re(\sqrt{s})$ is above the threshold of the
channel, the pole might correspond to a resonance state of vector
mesons and octet baryons. Otherwise, if $Re(\sqrt{s})$ of the pole
is less than the threshold, it is more possible to be a bound state.

When a pole of the amplitude $T^J_{j^\prime,j;l^\prime,l}$ is
produced in the complex plane of $\sqrt{s}$, not only the mass, the
decay width, the parity and the total angular momentum $J$ of the
corresponding resonance are determined, but the detailed information
on the orbital and total angular momenta $l,j$ and $l^\prime,
j^\prime$ of the initial and final vector mesons to generate this
resonance can also be obtained.

\section{Results}
\label{sect:Results}

In this section we show our results obtained in the channels of
strangeness zero and different isospins, respectively.

\subsection{Isospin I=1/2}

\begin{figure}[htb]
\begin{center}
\includegraphics[width=0.5\textwidth]{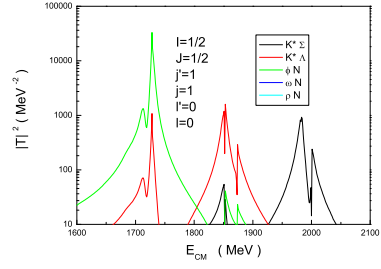}
\end{center}
\caption{$|T|^2$ for different channels with $I=1/2$, $J=1/2$,
$j^\prime=1$, $j=1$, $l^\prime=0$ and $l=0$.} \label{fig:J12-1100}
\end{figure}

\begin{figure}[htb]
\begin{center}
\includegraphics[width=0.5\textwidth]{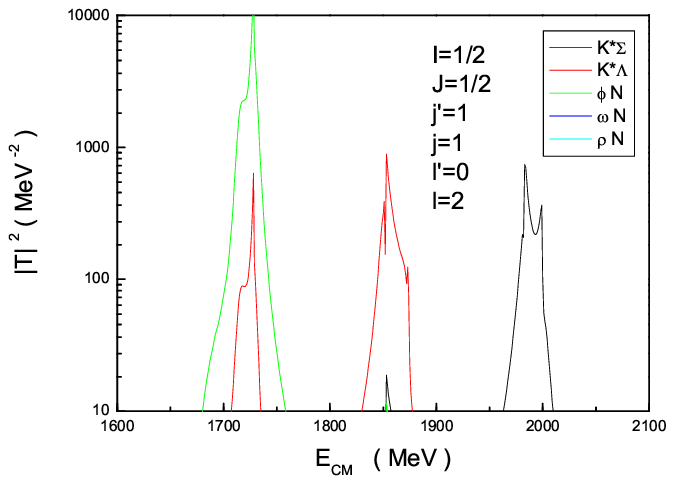}
\end{center}
\caption{$|T|^2$ for different channels with $I=1/2$, $J=1/2$,
$j^\prime=1$, $j=1$, $l^\prime=0$ and $l=2$.} \label{fig:J12-1102}
\end{figure}

\begin{figure}[htb]
\begin{center}
\includegraphics[width=0.5\textwidth]{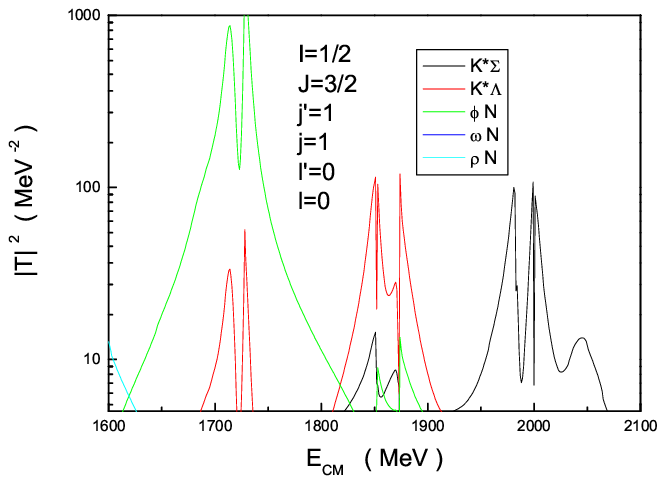}
\end{center}
\caption{$|T|^2$ for different channels with $I=1/2$, $J=3/2$,
$j^\prime=1$, $j=1$, $l^\prime=0$ and $l=0$.} \label{fig:J32-1100}
\end{figure}

\begin{figure}[htb]
\begin{center}
\includegraphics[width=0.5\textwidth]{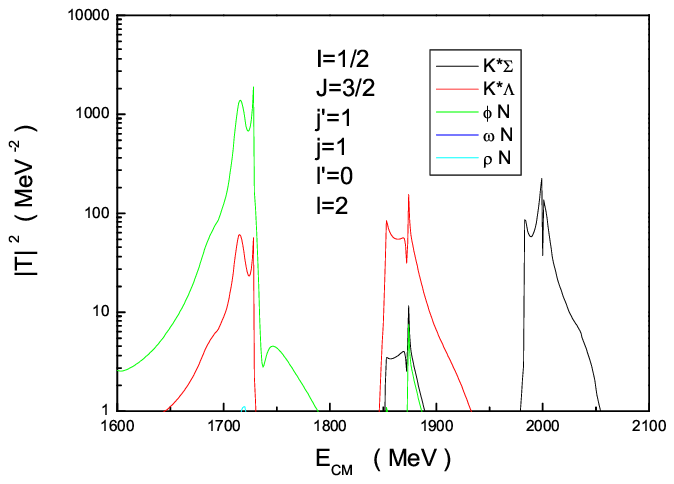}
\end{center}
\caption{$|T|^2$ for different channels with $I=1/2$, $J=3/2$,
$j^\prime=1$, $j=1$, $l^\prime=0$ and $l=2$.} \label{fig:J32-1102}
\end{figure}

\begin{figure}[htb]
\begin{center}
\includegraphics[width=0.5\textwidth]{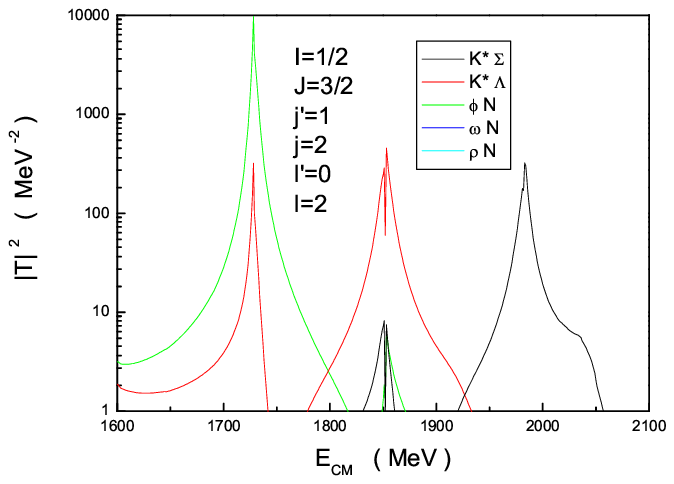}
\end{center}
\caption{$|T|^2$ for different channels with $I=1/2$, $J=3/2$,
$j^\prime=1$, $j=2$, $l^\prime=0$ and $l=2$.} \label{fig:J32-1202}
\end{figure}

In the channel of Isospin $I=1/2$, the isospin states for $\rho N$,
$\omega N$, $\phi N$, $K^* \Lambda$, $K^* \Sigma$ can be written as
\begin{equation}
\label{isospin12-rN} |\rho N; \frac{1}{2}, \frac{1}{2} \rangle
~=~-\sqrt{\frac{1}{3}}|\rho_0 p \rangle~-~\sqrt{\frac{2}{3}}|\rho_+
n \rangle,
\end{equation}

\begin{equation}
|\omega N; \frac{1}{2}, \frac{1}{2} \rangle~=~|\omega p\rangle,
\end{equation}

\begin{equation}
|\phi N; \frac{1}{2}, \frac{1}{2} \rangle~=~|\phi p \rangle,
\end{equation}

\begin{equation}
|K^* \Lambda; \frac{1}{2}, \frac{1}{2} \rangle~=~|{K^*}^+ \Lambda
\rangle,
\end{equation}
and
\begin{equation}
\label{isospin12-KS} |K^* \Sigma; \frac{1}{2}, \frac{1}{2}
\rangle~=~\sqrt{\frac{1}{3}}|{K^*}^+ \Sigma^0
\rangle~+~\sqrt{\frac{2}{3}}|{K^*}^0 \Sigma^+ \rangle,
\end{equation}
respectively. We have used the phase convention $\rho^+=-|1,1>$ and
$\Sigma^+=-|1,1>$ for the isospin states in
Eqs.~(\ref{isospin12-rN}) and (\ref{isospin12-KS}), which is
consistent with the structure of the $V^\mu$ and $B$ matrices. It is
apparent that the interaction between isospin states with isospin
orientation $I_z=-1/2$ would generate the same resonances, so only
the isospin states with $I_z=1/2$ are discussed in this section.

%
%
In Fig.~\ref{fig:J12-1100} we show the results of $|T_{ii}|^2$ as a
function of $\sqrt{s}$ for the different channels in the
$(I,J,j^\prime,j,l^\prime,l)=(1/2,1/2,1,1,0,0)$ sector. We can see
two peaks for $\phi N \rightarrow \phi N$ around 1700 MeV, a few MeV
above the $\rho N$ threshold. These two peaks around 1700 MeV are
also seen in the $K^* \Lambda$ channel but are absent or barely
visible in the $\rho N$, $\omega N$ and $K^* \Sigma$ channels.
On the other hand, with the same quantum numbers
$(I,J,j^\prime,j,l^\prime,l)$, one finds other two peaks around 1860
MeV. which are clearly visible in the $K^* \Lambda$, $K^* \Sigma$
and $\phi N$ channels but not visible in the $\rho N$ and $\omega N$
ones. Moreover, there are also two peaks appeared around 2000 MeV in
the $K^* \Sigma$ channel, which are not visible in the other
channels.
$|T_{ii}|^2$ as a function of $\sqrt{s}$ for the different channels
with $(I,J,j^\prime,j,l^\prime,l)=(1/2,1/2,1,1,0,2)$ is depicted in
Fig.~\ref{fig:J12-1102}. The peaks appear almost at the same
positions in the complex plane of $\sqrt{s}$ as those in
Fig.~\ref{fig:J12-1100}.

The couplings of these resonances to vector mesons and octet baryons
are different when the quantum numbers $(j^\prime, j, l^\prime, l)$
take different values. In follows, we only calculate their coupling
constants to different vector mesons and octet baryons in the
channel of $l=l^\prime=0$, i.e., $j=j^\prime=1$.
The couplings of the resonances to different channels for the
$(I,J,j^\prime,j,l^\prime,l)=(1/2,1/2,1,1,0,0)$ sector, obtained
from the residues at the poles are shown in
Table~\ref{coup-S0I12-J12}. It is apparent that the poles at
$1715+i4$~MeV and $1728+i0$~MeV couple strongly to the $\phi N$
channel, while the poles at $1855+i1$~MeV and $1868+i6$~MeV couple
strongly to the $K^* \Lambda$ channels. For the two poles around
2000 MeV, at the positions of $1982+i4$~MeV and $1999+i5$~MeV,
mainly interact with the $K^* \Sigma$ channel.

\begin{table}[htbp]
\begin{tabular}{c|c|c|c|c|c}
\hline\hline
 Pole positions   &  $\rho N$          &  $\omega N$   &   $\phi N$          & $K^* \Lambda$ & $K^* \Sigma$ \\
\hline
$1715+i4$~MeV & $-0.60-i0.28$  &  $0.43-i0.03$ &  $4.29-i0.05$ & $-1.98-i0.08$ & $0.18+0.03$  \\
$1728+i0$~MeV & $0.0-i0.63$    &  $0.0+i0.40$  &  $0.0+i7.42$  & $0.0-i3.03$   & $0.0+i0.37$  \\
$1855+i1$~MeV  &$-0.14-i0.14$   &  $-0.31-i0.32$ &  $1.52-i0.01$ & $4.50+i0.00$  & $1.85+i0.03$    \\
$1868+i6$~MeV  & $-0.13-i0.14$  & $-0.31-i0.37$ & $1.26-i0.07$  & $2.99-i0.06$ & $1.53+i0.02$ \\
$1982+i4$~MeV  &$0.10+i0.27$   &  $-0.03+i0.02$  & $-0.28-i0.19$  & $-1.01-i0.17$ & $5.62-i0.18$ \\
$1999+i5$~MeV & $0.07+i0.21$  &   $-0.02+i0.02$  &   $-0.20-i0.16$ &   $-0.59-0.12$   & $2.75-0.12$   \\
\hline \hline
\end{tabular}
\caption{Pole positions and coupling constants to various channels
of resonances found in the isospin $I=1/2$ and spin $J=1/2$ sector.}
\label{coup-S0I12-J12}
\end{table}

%
%

The case of $J=3/2$ also shows clear double peaks around 1700 MeV,
1860 MeV and 2000 MeV in the $(j^\prime, j, l^\prime, l)=(1,1,0,0)$
and $(j^\prime, j, l^\prime, l)=(1,1,0,2)$ sectors, which are
depicted in Figs.~\ref{fig:J32-1100}, and ~\ref{fig:J32-1102}
respectively. Similarly to the case of $J=1/2$, the double peaks
around 1700 MeV are visible in the $\phi N$ and $K^* \Lambda$
channels, and the double peaks around 1860 MeV appear mainly in the
$K^* \Lambda$, $\phi N$ and $K^* \Sigma$ channels. Moreover, in
Fig.~\ref{fig:J32-1100} a lower peak can be seen around 2050 MeV
near the double peaks around 2000 MeV in the $K^* \Sigma$ channel.
In Fig.~\ref{fig:J32-1202} the sector $(j^\prime, j, l^\prime,
l)=(1,2,0,2)$ is displayed, and we can see only one peak is left
around 1700 MeV, 1860 MeV and 2000 MeV, respectively.

\begin{table}[htbp]
\begin{tabular}{c|c|c|c|c|c}
\hline\hline
 Pole positions   &  $\rho N$          &  $\omega N$   &   $\phi N$          & $K^* \Lambda$ & $K^* \Sigma$ \\
\hline
$1715+i4$~MeV & $-0.60-i0.28$  & $0.43-i0.02$   & $4.29-i0.05$     & $-1.98-i0.08$ & $0.18+i0.03$  \\
$1868+i6$~MeV & $-0.13-i0.14$   & $-0.31-i0.37$ & $1.26-i0.07$ &     $2.99-i0.06$  &  $1.53+0.02$  \\
$1982+i4$~MeV  & $0.05+i0.13$  & $-0.01+i0.01$    & $-0.14-i0.10$  & $-0.51-i0.08$ & $2.81-i0.09$ \\
$1999+i5$~MeV &  $0.07+i0.21$ &  $-0.02+i0.02$ & $-0.19-i0.16$ & $-0.59-i0.12$ & $2.75-i0.12$   \\
$2045+i25$~MeV &  $0.02+i0.48$ & $-0.25+i0.10$  &$-0.21-i0.48$  & $-0.66-i0.59$ & $3.03-i0.46$      \\
\hline \hline
\end{tabular}
\caption{Pole positions and coupling constants to various channels
of resonances found in the isospin $I=1/2$ and spin $J=3/2$ sector.}
\label{coup-S0I12-J32}
\end{table}

The couplings of the resonances to different channels for the
$(I,J,j^\prime,j,l^\prime,l)=(1/2,3/2,1,1,0,0)$ sector are listed in
Table~\ref{coup-S0I12-J32}. Comparing with the
$(I,J,j^\prime,j,l^\prime,l)=(1/2,1/2,1,1,0,0)$ sector, we can see
both $J=3/2$ and $J=1/2$ resonances are generated dynamically at the
same positions of the complex plane of $\sqrt{s}$, i.e.,
$1715+i4$~MeV, $1868+i6$~MeV and $1982+i4$~MeV.
Furthermore, the couplings of some resonances to different channels
take the same values as those of the $J=1/2$ case, especially at the
positions of $1715+i4$~MeV and $1868+i6$~MeV.

The poles at $1982+i4$~MeV, $1999+i5$~MeV and $2045+i25$~MeV all
couple strongly to the $K^* \Sigma$ channel, and their couplings to
different channels are similar to each other. It implies that these
three poles might correspond to one resonance.

\begin{table}[htbp]
\begin{tabular}{c|c|c|c|c|c}
\hline\hline
 $J$ & Theory  &PDG data \\
\hline
 & Pole positions   &  Name and $J^P$          &   Status     &   Mass          &  Width \\
\hline
1/2 &$1715+i4$~MeV &  $N(1650)1/2^{-}$& **** &$1645-1670$~MeV &$120-180$~MeV \\
3/2 &$1715+i4$~MeV &  $N(1700)3/2^-$  & ***     &    $1650-1750$~MeV  &   $100-250$~MeV \\
1/2 &$1728+i0$~MeV  &                  &          &                     &    \\
1/2 &$1855+i1$~MeV  &                  &          &                     &    \\
1/2 &$1868+i6$~MeV  &  $N(1895)1/2^{-}$ &  **    &  $\approx 2090$~MeV & $100-400$~MeV    \\
3/2&$$1868+i6$$~MeV &  $N(1875)3/2^-$  & *** &$1820-1920$~MeV  &  $160-320$~MeV             \\
1/2 &$1982+i4$~MeV  &   &   &  &       \\
3/2& $1982+i4$~MeV & $N(2120)3/2^-$  & **  & $\approx2120$~MeV & $\approx300$~MeV      \\
1/2&$1999+i5$~MeV & &  & &    \\
3/2&$1999+i5$~MeV & &  & &    \\
3/2&$2045+i25$~MeV & &  & &    \\
\hline \hline
\end{tabular}
\caption{The properties of the dynamically generated resonances with
isospin $I=1/2$ and their possible PDG counterparts.}
\label{table:i12}
\end{table}

In Table~\ref{table:i12} we show a summary of the results obtained
and the tentative association to known states.

For the $(I,J)=(1/2,3/2)$ $N^*$ states there is the $N(1700)$ with
$J^P=3/2^-$, which could be associated with the state we find with
the same quantum number at $1715+i4$~MeV.
There are two resonances $N(1700)3/2^-$ and $N(1685)?^?$ in the same
energy region of PDG data, while the total angular momentum $J$ and
parity of the latter are not determined. Since the state
$N(1685)?^?$ does not gain status by being a sought-after member of
a baryon anti-decuplet, we tend to assume the pole appeared at
$1715+i4$~MeV might be the resonance $N(1700)3/2^-$.
We also find the resonance at $1715+i4$~MeV in the $J^P=1/2^-$
sector, and this $J^P=1/2^-$ states could correspond to the
$N(1650)1/2^{-}$, which could be the spin partner of the
$N(1700)3/2^-$.
Our calculation shows the $N(1650)1/2^{-}$ couples strongly to the
$\phi N$ and $K^* \Lambda$ channels. At this point, it is different
from the results obtained in Ref.~\cite{ramos2010}, where the
coupling constant to the channel $\rho N$ is the largest.

In the $J=3/2$ sector the poles at $1868+i6$~MeV and $1982+i4$~MeV
could correspond to the resonances $N(1875)3/2^-$ and
$N(2120)3/2^-$, respectively. These two resonances had been labeled
as one resonance $N(2080)$ before the 2012 PDG Review\cite{pdg}.

In the region above 1800~MeV, only one resonance $N(1895)$ is listed
with $J^P=1/2^{-}$ in the PDG data, which appears in the PDG review
as $N^*(2090)~S_{11}$ before 2012\cite{pdg}.
Although an estimated mass value of the state $N(1895)1/2^{-}$ about
2090 MeV is listed in the PDG review\cite{pdg}, the newest
multichannel analysis manifests the mass of this state is 1895 MeV
in Ref.~\cite{Anisovich}, which is close to the mass of the state
$N(1875)3/2^-$. Thus we treat the $N(1895)1/2^{-}$ as a spin partner
of the $N(1875)3/2^-$, and assume the pole at $1868+i6$~MeV in the
$J=1/2$ sector could correspond to the $N(1895)1/2^{-}$.

In addition, a high peak is also found at $1982+i4$~MeV in the
$J=1/2$ sector, and no counterpart is listed in the PDG data. It
might stimulate the experimental research to look for the resonance
around 2000 MeV, which might be a spin partner of the $N(2120)3/2^-$
and couple strongly to the $K^* \Sigma$ channel as listed in
Table~\ref{coup-S0I12-J12}.

\subsection{Isospin I=3/2}

The isospin wave functions for the $I=3/2$ case are written as
\begin{equation}
|\rho N; \frac{3}{2}, \frac{1}{2}\rangle~=~\sqrt{\frac{2}{3}}|\rho_0
p \rangle~-~\sqrt{\frac{1}{3}}|\rho_+ n \rangle,
\end{equation}
and
\begin{equation}
|K^* \Sigma; \frac{3}{2}, \frac{1}{2}
\rangle~=~\sqrt{\frac{2}{3}}|{K^*}^+ \Sigma^0
\rangle~-~\sqrt{\frac{1}{3}}|{K^*}^0 \Sigma^+ \rangle,
\end{equation}
respectively.
The coupling constants of octet baryons and vector mesons take the
same values when the isospin orientation $I_z$ is different from one
another. Especially, the $s-$ channel coupling constants tend to
zero, and it is apparent that the $s-$ channel interaction is
forbidden in the case of isospin $I=3/2$.

\begin{figure}[htb]
\begin{center}
\includegraphics[width=0.5\textwidth]{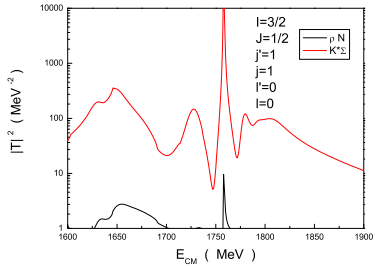}
\end{center}
\caption{$|T|^2$ for different channels with $I=3/2$, $J=1/2$,
$j^\prime=1$, $j=1$, $l^\prime=0$ and $l=0$.}
\label{fig:J12-1100-i32}
\end{figure}

\begin{figure}[htb]
\begin{center}
\includegraphics[width=0.5\textwidth]{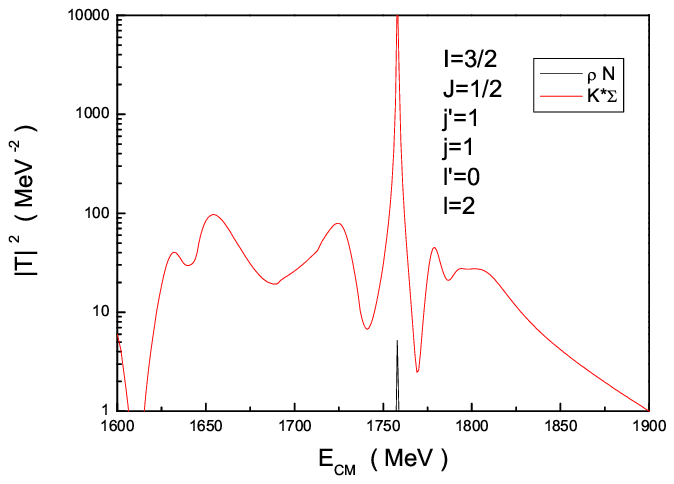}
\end{center}
\caption{$|T|^2$ for different channels with $I=3/2$, $J=1/2$,
$j^\prime=1$, $j=1$, $l^\prime=0$ and $l=2$.}
\label{fig:J12-1102-i32}
\end{figure}

\begin{figure}[htb]
\begin{center}
\includegraphics[width=0.5\textwidth]{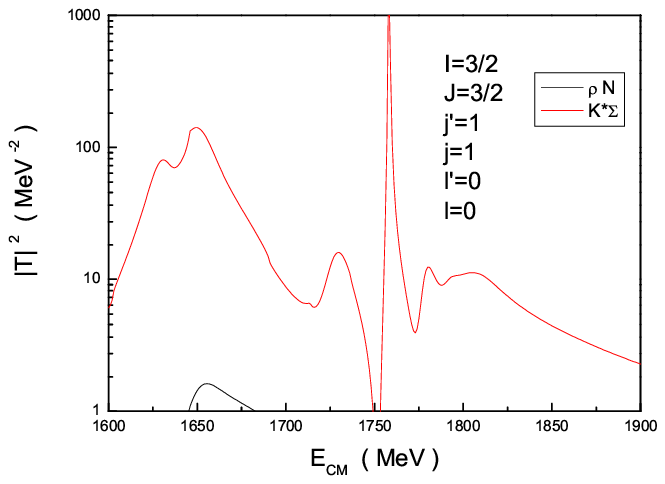}
\end{center}
\caption{$|T|^2$ for different channels with $I=3/2$, $J=3/2$,
$j^\prime=1$, $j=1$, $l^\prime=0$ and $l=0$.}
\label{fig:J32-1100-i32}
\end{figure}

\begin{figure}[htb]
\begin{center}
\includegraphics[width=0.5\textwidth]{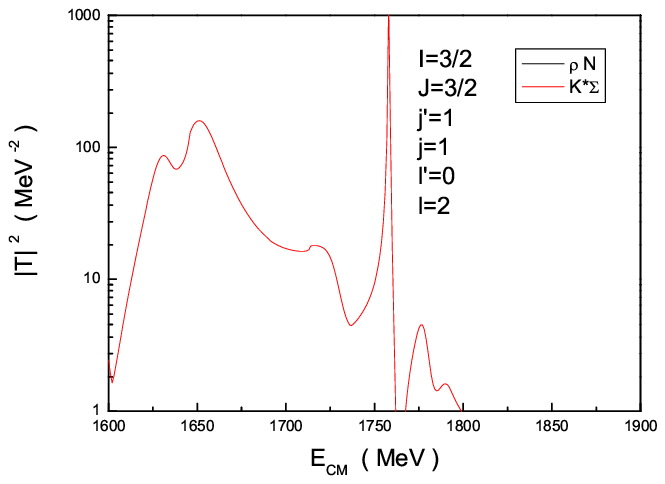}
\end{center}
\caption{$|T|^2$ for different channels with $I=3/2$, $J=3/2$,
$j^\prime=1$, $j=1$, $l^\prime=0$ and $l=2$.}
\label{fig:J32-1102-i32}
\end{figure}

\begin{figure}[htb]
\begin{center}
\includegraphics[width=0.5\textwidth]{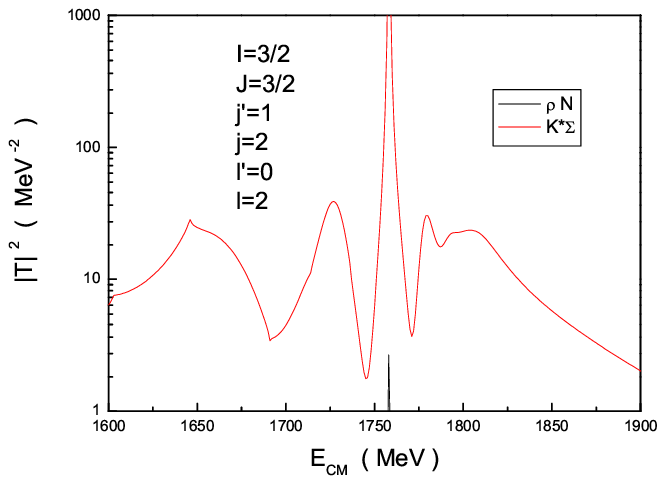}
\end{center}
\caption{$|T|^2$ for different channels with $I=3/2$, $J=3/2$,
$j^\prime=1$, $j=2$, $l^\prime=0$ and $l=2$.}
\label{fig:J32-1202-i32}
\end{figure}

\begin{table}[htbp]
\begin{tabular}{c|c|c}
\hline\hline
 Pole positions   &  $\rho N$         & $K^* \Sigma$ \\
\hline
$1619+i160$~MeV & $0.87-i2.64$   & $4.93-i4.95$   \\
$1654+i11$~MeV  & $3.04+i0.40$   &  $12.35+i0.04$  \\
$1680+i163$~MeV &  $-1.14-i0.87$  & $4.09-i3.72$    \\
$1726+i9$~MeV   &  $1.34+i0.41$  &  $8.89+i0.01$  \\
$1758+i0$~MeV   &  $0.00+i0.57$  &  $0.00+i2.08$  \\
$1780+i14$~MeV  & $0.68+i0.27$   &  $6.45+i0.05$   \\
\hline \hline
\end{tabular}
\caption{Pole positions and coupling constants to various channels
of resonances found in the isospin $I=3/2$ and spin $J=1/2$ sector.}
\label{coup-S0I32-J12}
\end{table}

\begin{table}[htbp]
\begin{tabular}{c|c|c}
\hline\hline
 Pole positions   &  $\rho N$         & $K^* \Sigma$ \\
\hline
$1619+i160$~MeV & $0.43-i1.32$    & $2.47-i2.48$   \\
$1654+i11$~MeV  & $1.52+i0.20$    & $6.17+i0.01$   \\
$1726+i9$~MeV   & $0.67+i0.21$    & $4.45+i0.01$   \\
$1758+i0$~MeV   & $0.00+i0.44$    & $0.00+i1.11$   \\
\hline \hline
\end{tabular}
\caption{Pole positions and coupling constants to various channels
of resonances found in the isospin $I=3/2$ and spin $J=3/2$ sector.}
\label{coup-S0I32-J32}
\end{table}

We show the $|T_{ii}|^2$ as a function of $\sqrt{s}$ for different
partial waves in Figs.~\ref{fig:J12-1100-i32},
~\ref{fig:J12-1102-i32}, ~\ref{fig:J32-1100-i32},
~\ref{fig:J32-1102-i32} and ~\ref{fig:J32-1202-i32}, respectively,
and we find four peaks in these figures can be seen, which
correspond to four poles near the real axis in the complex plane
$\sqrt{s}$, i.e., $1654+i11$~MeV, $1726+i9$~MeV, $1758+i0$~MeV and
$1780+i14$~MeV. Moreover, there are two other poles at
$1619+i160$~MeV and $1680+i163$~MeV, which are far from the real
axis and do not show clearly in the figures. All these states couple
strongly to the $K^* \Sigma$ channel. The couplings of these
resonances to the $\rho N$ and $K^* \Sigma$ channels are listed in
Table~\ref{coup-S0I32-J12} for the $J=1/2$ case and
Table~\ref{coup-S0I32-J32} for the $J=3/2$ case.

For the case of $(I,J)=(3/2,1/2)$ there is one state in the PDG, the
$\Delta(1620)1/2^-$, which has the spin partner $\Delta(1700)3/2^-$
with $J^P=3/2^-$. These two states could be associated with the pole
at $1654+i11$~MeV, which appeared both in the $J=1/2$ and $J=3/2$
sectors by the method of partial wave analysis in
Eq.~(\ref{eq:pwa}).

We find a narrow peak at $1758+i0$~MeV in the $J=1/2$ and $J=3/2$
sectors, and its couplings to $\rho N$ and $K^* \Sigma$ are smaller
by far than those of other states. We suspect it could be a cusp.

In Ref.~\cite{ramos2010}, no resonance is found in the channel of
isospin $I=3/2$ and strangeness zero since only $t-$ channel is
taken into account, which supply a repulsive interaction between
vector mesons and octet baryons.
However, our calculation results manifest some resonances can be
produced dynamically in the $I=3/2$ sector when we take into account
the other interaction modes besides $t-$ channel, especially the
contact term between vector mesons and octet baryons. At this point,
our results are consistent to those in Ref.~\cite{Hosaka2011}.

The states found in the $I=3/2$ sector are summarized in
Table~\ref{table:i32}, where the properties of the possible
counterparts are also listed. Except the states at $1654+i11$~MeV
with $J=1/2$ and $J=3/2$, there are not PDG counterparts associated
with the other states.

\begin{table}[htbp]
\begin{tabular}{c|c|c|c|c|c}
\hline\hline
 $J$ & Theory & PDG data \\
\hline
 & Pole positions                &  Name and $J^P$          &   Status     &   Mass          &  Width \\
\hline
 $1/2$ & $1619+i160$~MeV  &  &   &   &  \\
 $3/2$ & $1619+i160$~MeV  &  &   &   &  \\
 $1/2$ & $1654+i11$~MeV &$\Delta(1620)1/2^-$  &    ****  & $1600-1660$~MeV   & $130-150$~MeV \\
 $3/2$  &$1654+i11$~MeV &$\Delta(1700)3/2^-$  & ****     & $1670-1750$~MeV &  $200-400$~MeV  \\
 $1/2$ & $1680+i163$~MeV  &  &   &   &  \\
 $1/2$ & $1726+i9$~MeV  &  &   &   &  \\
 $3/2$ & $1726+i9$~MeV  &  &   &   &  \\
 $1/2$ & $1758+i0$~MeV  &  &   &   &  \\
 $3/2$ & $1758+i0$~MeV  &  &   &   &  \\
 $1/2$ & $1780+i14$~MeV  &  &   &   &  \\
\hline \hline
\end{tabular}
\caption{The properties of dynamically generated resonances with
isospin $I=3/2$ and its possible PDG counterparts.}
\label{table:i32}
\end{table}

\section{Summary}
\label{sect:Summary}

We have studied the interaction between vector mesons and octet
baryons using a unitary framework in coupled channels. In addition
to the vector interaction term, a tensor term is included in the
Lagrangian of vector mesons and octet baryons. With this interaction
Lagrangian, we obtain the kernels between vector mesons and octet
baryons from a summation of diagrams corresponding to a vector meson
exchange in the $t-$ channel, octet baryon exchange in $s-$, $u-$
channels, and a contact interaction related only to the tensor term.
The scattering amplitudes are calculated by solving the
coupled-channel Lippman-Schwinger equations, and then the results
are studied with the method of partial wave analysis.

In the $I=1/2$ sector, the double-peak structure of $|T_{ii}|^2$ is
found around 1700 MeV, 1860 MeV and 2000 MeV respectively in the
different partial waves, and these double-peaks could correspond to
the states in the PDG data.
The pole found at $1715+i4$~MeV with different $J$ in the complex
energy plane might correspond to the states $N(1650)1/2^{-}$ in the
$J=1/2$ case, and its spin partner $N(1700)3/2^-$ in the $J=3/2$
case. Similarly, the pole at $1868+i6$~MeV could correspond to
another pair of spin partners, $N(1895)1/2^{-}$ and $N(1875)3/2^-$
of the PDG data. The $N(2120)3/2^-$ could be associated with the
pole at $1982+i4$~MeV in the $J=3/2$ case. However, the spin partner
of $N(2120)3/2^-$ is absent in the PDG data, and we predict that
there should be a state around 2000 MeV in the $J=1/2$ case, which
should be associated with the pole at $1982+i4$~MeV with $J=1/2$ as
a spin partner of the state $N(2120)3/2^-$.

In the $I=3/2$ sector, we also find some poles in the complex energy
plane, and we assume the spin partners $\Delta(1620)1/2^-$ and
$\Delta(1700)3/2^-$ could be associated with the pole at
$1654+i11$~MeV in the $J=1/2$ and $J=3/2$ cases.

Furthermore, it manifests that the interaction between vector mesons
and octet baryons in the $I=3/2$ case can be attractive when the
$u-$, $s-$ channels and the contact term are taken into account
besides $t-$ channel interaction.
In addition, there are not PDG counterparts for some resonances in
our prediction, and it might give a stimulus to search
experimentally for these resonance states.

\section*{Acknowledgments}

Bao-Xi Sun thanks the useful discussion with E. Oset, Li-Sheng Geng
and Hua-Xing Chen. This work is supported by the National Natural
Science Foundation of China under grant number 10775012.

\section*{Appendix I: Vertices used in the calculation}

\begin{figure}[htb]
\begin{center}
\includegraphics[width=0.7\textwidth]{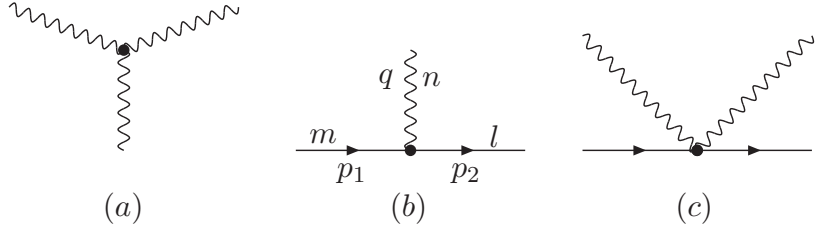}
\end{center}
\caption{Vertices used in the calculation. $(a)$ Three-vector-meson
vertex, $(b)$ Baryon-baryon-meson vertex, where the momentum of the
vector meson $q=p_2-p_1$. $(c)$ Contact vertex of octet baryons and
vector mesons.} \label{fig:vertex}
\end{figure}

\subsection{Vertices for three vector mesons}

The vertex for three vector mesons can be deduced from the
Lagrangian
\begin{eqnarray}
\mathcal{L}_{3V}&=&\frac{ig}{2}\sum_{l,m,n=1}^{8}C_{3V}(l,m,n)M_l^\mu M_m^\nu \left( \partial_\mu {M_n}_\nu~-~\partial_\nu {M_n}_\mu \right)
\end{eqnarray}
with
\begin{equation}
\label{coef:3v}
C_{3V}(l,m,n)~=~\sum_{i,j,k=1}^{8}X_{il} X_{jm} X_{kn} \left( d_{ijk} + i f_{ijk} \right).
\end{equation}
In Eq.~(\ref{coef:3v}), the matrix $X$ is indicated as
\begin{eqnarray}
X =\frac{1}{\sqrt{2}}
\left( \begin{array}{cccccccc}
1  & 1 & 0 & 0  & 0 & 0 & 0 & 0 \\
i & -i & 0 & 0  & 0 & 0 & 0 & 0 \\
0 & 0 & \sqrt{2} & 0  & 0 & 0 & 0 & 0 \\
0 & 0 & 0 & 1  & 1 & 0 & 0 & 0 \\
0 & 0 & 0 & i  & -i & 0 & 0 & 0 \\
0 & 0 & 0 & 0  & 0 & 1 & 1 & 0 \\
0 & 0 & 0 & 0  & 0 & i & -i & 0 \\
0 & 0 & 0 & 0  & 0 & 0 & 0 & \sqrt{2} \\
\end{array}\right),
\end{eqnarray}
and the values of $d_{ijk}$ and $f_{ijk}$ can be found in
Ref.~\cite{pdg}. Moreover, the coefficient in Eq.~(\ref{coef:3v})
must be multiplied by a factor of $\frac{1}{\sqrt{3}}$ for each
$\omega$ meson line, and a factor of $-\sqrt{\frac{2}{3}}$ for each
$\phi$ meson line. The vertices for three vector mesons are depicted
in Fig.~\ref{fig:vertex}$(a)$.

\subsection{Baryon-baryon-meson Vertices}

The interaction vertices for two baryons and one vector meson
depicted in Fig.~\ref{fig:vertex}$(b)$ can be obtained according to
$SU(3)$ symmetry
\begin{equation}
\Gamma^\mu(p_2,p_1)~=~g_1 \gamma^\mu~+~g_2 (p_2^\mu~+~p_1^\mu),
\end{equation}
where
\begin{eqnarray}
g_1(l,m,n)&=&g\left((F_V + D_V) C_{\bar{B}BV}(l,n,m)+( - F_V + D_V ) C_{\bar{B}BV}(l,m,n) \right)  \\
&+&\frac{g}{2M_N}\left((F_T + D_T) C_{\bar{B}BV}(l,n,m)+( - F_T +
D_T ) C_{\bar{B}BV}(l,m,n) \right) (M_l+M_m), \nonumber
\end{eqnarray}
and
\begin{eqnarray}
g_2(l,m,n)&=&-\frac{g}{2M_N}
    \left((F_T + D_T) C_{\bar{B}BV}(l,n,m)+( - F_T + D_T ) C_{\bar{B}BV}(l,m,n) \right)
\end{eqnarray}
with
\begin{equation}
C_{\bar{B}BV}(l,m,n)=  \frac{1}{2} \sum_{i,j,k=1}^{8} X^\dagger_{li}
X_{jm} X_{kn} \left( d_{ijk} + i f_{ijk} \right).
\end{equation}

\subsection{Contact term}

From the lagrangian in Eq.~(\ref{contactl}), the coupling constant of the contact term is written as
\begin{eqnarray}
C(i^\prime, j^\prime, k^\prime, l^\prime)=\frac{ig^2}{32M_N}\sum_{i,j,k,l=1}^{8}&&\{F_T
[ \langle  \lambda_i \lambda_j \lambda_k \lambda_l \rangle
- \langle  \lambda_i \lambda_k \lambda_j \lambda_l \rangle
- \langle  \lambda_i \lambda_l \lambda_j \lambda_k \rangle
+ \langle  \lambda_i \lambda_l \lambda_k \lambda_j \rangle
] \nonumber \\
&+&D_T  [ \langle  \lambda_i \lambda_j \lambda_k \lambda_l \rangle
- \langle  \lambda_i \lambda_k \lambda_j \lambda_l \rangle
+ \langle  \lambda_i \lambda_l \lambda_j \lambda_k \rangle
- \langle  \lambda_i \lambda_l \lambda_k \lambda_j \rangle
]
\} \nonumber \\
&&X^\dagger_{i^\prime,i} X_{j,j^\prime} X_{k,k^\prime} X_{l,l^\prime},
\end{eqnarray}
where
\begin{equation}
\langle  \lambda_i \lambda_j \lambda_k \lambda_l \rangle = \frac{4}{3}\delta_{ij}\delta_{kl} + 2\left(d_{klm}+if_{klm} \right) \left(d_{ijm}+if_{ijm} \right)
\end{equation}
with $\lambda_i$ the hermitian generator of group $SU(3)$\cite{pdg}.
The vertex of contact terms is depicted in Fig.~\ref{fig:vertex}$(c)$.
We can obtain the coupling constants $C^{CT}_{IS}$ of the contact
term for different isospins with Clebsch-Gordan coefficients.

\section*{Appendix II: Polarization vectors and Dirac spinors}

\subsection{Polarization vectors of vector mesons}

The polarization vector of the massive vector field satisfies the
constraint
\begin{equation}
\label{eq:kcdotepsilon} k \cdot \varepsilon_{\vec{k}\lambda}~=~0,
\end{equation}
with $k$ the momentum of the vector meson.

For each $k$, a set of three linearly independent polarization
vectors satisfying Eq.~(\ref{eq:kcdotepsilon}) can be constructed as
in Ref.~\cite{Lurie}. If $\vec{\varepsilon}_{\vec{k}\lambda}$
$(\lambda~=~1,2,3)$ is any triad of three-vectors satisfying the
orthonormality relations
\begin{equation}
\vec{\varepsilon}_{\vec{k}\lambda} \cdot
\vec{\varepsilon}^*_{\vec{k}\lambda^\prime}~=~\delta_{\lambda
\lambda^\prime},
\end{equation}
then the three four-vectors can be written as
\begin{equation}
\label{eq:epsilon} \varepsilon_{\vec{k}\lambda}^\alpha~=~\left\{
\begin{array}{c}
\frac{\left(\vec{k}\cdot\vec{\varepsilon}_{\vec{k}\lambda}
\right)}{\mu},~~~~(\alpha~=~0), \\
\\
\vec{\varepsilon}_{\vec{k}\lambda}~+~\frac{\vec{k}\left(\vec{k}\cdot\vec{\varepsilon}_{\vec{k}\lambda}
\right)} {\mu \left( \omega_k~+~\mu \right)}, ~~~~(\alpha~=~1,2,3),
\end{array}
\right.
\end{equation}
with $\mu$ the mass and $\omega_k=\sqrt{\vec{k}^2+\mu^2}$ the energy
of the vector meson.
The four-vectors in Eq.~(\ref{eq:epsilon}) satisfy both
Eq.~(\ref{eq:kcdotepsilon}) and the orthonormality relations
\begin{equation}
\label{eq:orthonormality} \varepsilon_{\vec{k}\lambda} \cdot
\varepsilon^*_{\vec{k}\lambda^\prime}~=~-g_{\lambda \lambda^\prime},
\end{equation}
with the metric tensor
\begin{eqnarray}
g_{\lambda \lambda^\prime} = \left( \begin{array}{cccc}
1 & 0 & 0 & 0 \\
0 & -1 & 0 & 0\\
0 &0 & -1 & 0 \\
0 &0 & 0 & -1 \\
\end{array}\right).
\end{eqnarray}
Therefore, the notations used in this article are different from
those in Ref.~\cite{Lurie}.

A convenient choice for the triad of orthogonal unit vectors
$\vec{\varepsilon}_{\vec{k}\lambda}$ is to take
$\vec{\varepsilon}_{\vec{k}3}$ pointing along the three-momentum
$\vec{k}$ with $\vec{\varepsilon}_{\vec{k}1}$ and
$\vec{\varepsilon}_{\vec{k} 2}$ orthogonal both to
$\vec{\varepsilon}_{\vec{k}3}$ and to each other, i.e.
\begin{equation}
\label{eq:e3} \vec{\varepsilon}_{\vec{k}
3}~=~\frac{\vec{k}}{|\vec{k}|},
\end{equation}
and
\begin{equation}
\label{eq:e1e2} \vec{\varepsilon}_{\vec{k} 1} \cdot
\vec{k}~=~\vec{\varepsilon}_{\vec{k} 2} \cdot
\vec{k}~=~\vec{\varepsilon}_{\vec{k} 1} \cdot
\vec{\varepsilon}_{\vec{k} 2}~=~0. \nonumber
\end{equation}
The polarization vectors $\vec{\varepsilon}_{\vec{k} 1}$ and
$\vec{\varepsilon}_{\vec{k} 2}$ represent states of $transverse$
polarization, while $\vec{\varepsilon}_{\vec{k}3}$ represents
$longitudinal$ polarization.

In the scattering process of the vector meson and octet baryon, we
can choose the $transverse$ polarization vectors of the initial
vector meson to be
\begin{eqnarray}
\label{3polar-e1-e2} \vec{\varepsilon}_{\vec{k} 1}& =&
\frac{1}{\sqrt{2}} \left( \begin{array}{c}
1 \\
i \\
0 \\
\end{array}\right),~~~~
\vec{\varepsilon}_{\vec{k} 2} = \frac{1}{\sqrt{2}} \left(
\begin{array}{c}
1 \\
-i \\
0 \\
\end{array}\right),
\end{eqnarray}
and the $longitudinal$ polarization vector
\begin{eqnarray}
\label{3polar-e3} \vec{\varepsilon}_{\vec{k} 3}& =& \left(
\begin{array}{c}
0 \\
0 \\
1 \\
\end{array}\right),
\end{eqnarray}
with $\vec{k}$ the three-momentum of the initial vector meson. For
the final vector meson, the $transverse$ and $longitudinal$
polarization vectors take the same formula as those of the initial
vector meson in Eqs.~(\ref{3polar-e1-e2}) and (\ref{3polar-e3}).

The polarization vectors $\varepsilon_{\vec{k} 1}$,
$\varepsilon_{\vec{k} 2}$, $\varepsilon_{\vec{k} 3}$ and $k/m$ form
a quartet of orthonormal four-vectors. From the completeness
relation for this quartet we deduce that
\begin{equation}
\sum_{\lambda=1,2,3} \varepsilon^\mu_{\vec{k} \lambda}
{\varepsilon^\ast}^\nu_{\vec{k} \lambda}~=~-g^{\mu\nu}+\frac{k^\mu
{k^\ast}^\nu}{\mu^2}.
\end{equation}

\subsection{Dirac spin wave function of octet baryons}

The Dirac equation implies
\begin{equation}
\left( \rlap{/}p-M \right)U(p,\lambda)=0,
\end{equation}
where the spinor of octet baryons can be written as
\begin{equation}
U(p,\lambda)=\frac{ \rlap{/}p+M
}{\sqrt{2M(M+E)}}U(M,\vec{0},\lambda),
\end{equation}
with
\begin{eqnarray}
 U(M,\vec{0},1)& =&
\left( \begin{array}{c}
1 \\
0 \\
0 \\
0 \\
\end{array}\right)
\end{eqnarray}
and
\begin{eqnarray}
 U(M,\vec{0},2)& =&
\left( \begin{array}{c}
0 \\
1 \\
0 \\
0 \\
\end{array}\right).
\end{eqnarray}

The conjugate spinor of octet baryons is obtained as
\begin{equation}
\bar{U}(p,\lambda)=\bar{U}(M,\vec{0},\lambda)\frac{ \rlap{/}p+M
}{\sqrt{2M(M+E)}},
\end{equation}
and the normalization factors have been chosen in order that
\begin{equation}
\bar{U}(p,\lambda)U(p,\lambda^\prime)=\delta_{\lambda,\lambda^\prime}.
\end{equation}

\end{document}